\documentclass[aps,prb,reprint,superscriptaddress]{revtex4-1}

\usepackage{amsmath}
\usepackage{amssymb}
\usepackage{amsfonts}
\usepackage{amssymb}
\usepackage{xspace}
\usepackage{graphics} 
\usepackage[linktoc=all]{hyperref}
\usepackage{siunitx}
\usepackage[export]{adjustbox}
\usepackage{natbib}
\usepackage{bibentry}
\usepackage{datatool}
\usepackage{notes2bib}
\usepackage{patchcmd}

\DTLnewdb{bibnotes}
\def\bibnote#1#2{%
  \DTLnewrow{bibnotes}
  \DTLnewdbentry{bibnotes}{mylabel}{#1}
  \DTLnewdbentry{bibnotes}{mynote}{#2}
}

\newcommand{\imchi}{\ensuremath{\chi''}\xspace}

\newcommand{\omegakink}{\ensuremath{\Omega_{\text{kink}}}\xspace}

\let \vaccent=\v 
\renewcommand{\v}[1]{\ensuremath{\boldsymbol{#1}}} 

\renewcommand*{\Re}{\operatorname{Re}}
\renewcommand*{\Im}{\operatorname{Im}}

\bibnotesetup{
  note-name    = ,
  use-sort-key = false
}

\begin{document}


\title{Role of the upper branch of the hour-glass magnetic spectrum in the formation
of the main kink in the electronic dispersion of high-T$_\text{c}$ cuprate
superconductors}


\author{Dominique Geffroy}
\email{geffroy@mail.muni.cz}
\affiliation{Department of Condensed Matter Physics,
Faculty of Science, Masaryk University,
Kotl{\'a}{\vaccent{r}}sk{\'a} 2, 611 37 Brno, Czech Republic}

\author{Ji\vaccent{r}\'{\i} Chaloupka}
\affiliation{Department of Condensed Matter Physics,
Faculty of Science,
Masaryk University,
Kotl{\'a}{\vaccent{r}}sk{\'a} 2, 611 37 Brno, Czech Republic}
\affiliation{Central European Institute of Technology, Masaryk University, Kamenice
753/5, 62500 Brno, Czech Republic}

\author{Thomas Dahm}
\affiliation{Universit{\"a}t Bielefeld, Fakult{\"a}t f{\"u}r Physik, Postfach
  100131, D-33501 Bielefeld, Germany}

\author{Dominik Munzar}
\email{munzar@physics.muni.cz}
\affiliation{Department of Condensed Matter Physics,
Faculty of Science,
Masaryk University,
Kotl{\'a}{\vaccent{r}}sk{\'a} 2, 611 37 Brno, Czech Republic}
\affiliation{Central European Institute of Technology, Masaryk University, Kamenice
753/5, 62500 Brno, Czech Republic}


\date{\today}

\begin{abstract}
We investigate the electronic
dispersion of the high-T$_{\mathrm{c}}$ cuprate superconductors using
the fully self-consistent version of the phenomenological model, where
charge planar quasiparticles are coupled to spin fluctuations. The
inputs we use ---the underlying (bare) band structure and the spin
susceptibility $\chi$--- are extracted from fits of angle resolved
photoemission and inelastic neutron scattering data of underdoped
YBa$_{2}$Cu$_{3}$O$_{6.6}$ by T.~Dahm and coworkers (T.~Dahm {\it et
  al.}, Nat.~Phys.~{\bf 5}, 217 (2009)). Our main results are: (i) We
have confirmed the finding by T.~Dahm and coworkers that the main nodal kink is, for the present values
of the input parameters, determined by the upper branch of the
hour-glass of $\chi$. We demonstrate that the properties of the kink
depend qualitatively on the strength of the charge-spin 
coupling. (ii) The effect of the resonance mode of $\chi$ on the
electronic dispersion strongly depends on its
kurtosis in the quasimomentum space. A low (high) kurtosis implies a
negligible (considerable) effect of the mode on the dispersion in the
near-nodal region. (iii) The energy of the kink decreases as a
function of the angle $\theta$ between the Fermi surface cut and the
nodal direction, in qualitative agreement with recent experimental
observations. We clarify the trend and make a specific prediction
concerning the angular dependence of the kink energy in underdoped
YBa$_{2}$Cu$_{3}$O$_{6.6}$.

\end{abstract}

\maketitle

\section{Introduction}

The kink at \numrange[range-phrase =
--]{50}{80} \si{\milli\electronvolt} in the electronic dispersion along
the Brillouin zone diagonal (i.e., from $(0,0)$ to $(\pi,\pi)$) of
high-T$_\text{c}$ cuprate superconductors\cite{Valla1999a,
  Bogdanov2000, Kaminski2001, Lanzara2001a, Gromko2003, Sato2003,
  Zhang2008, Damascelli2003} has been the object of
intense scrutiny by the scientific community since it was
first reported. Understanding
of the kink may be of importance in the context of the quest for the mechanism of high temperature
superconductivity. Unfortunately, a satisfactory understanding has not
yet been achieved. While there is a broad (yet not
unanimous\cite{Devereaux2004, Cuk2004a, Anderson2007, Byczuk2007,
  Park2013a}) consensus that the kink is due to an interaction with
bosonic excitations, the nature of the latter excitations remains
controversial. It is debated whether they are of
lattice\cite{Lanzara2001a, Gweon2004, Lee2006, Iwasawa2008, Giustino2008,
  Reznik2008, Capone2010, Lanzara2010, Vishik2010} (phonon),
magnetic\cite{Norman1997a, Johnson2001b, Kaminski2001, Eschrig2002,
  Chubukov2004, Eschrig2003, Manske2004, Eschrig2006, Kordyuk2006,
  Borisenko2006, Zabolotnyy2007a, Zabolotnyy2007, Borisenko2006a, Das2012, Das2014,
  Kordyuk2006b} (spin fluctuation), or more complex\cite{Yun2011,
  Zhang2012, Hong2013, Hong2013a, He2013, Mazza2013}
origin\cite{Carbotte2011}.

Regarding the magnetic scenario, it has been claimed for some time
that the kink reflects the coupling of the charged quasiparticles to
the resonance mode observed by neutron
scattering\cite{Rossat-Mignod1991, Fong1996, Fong1999a, Fong2000}. In
a more recent study by Dahm and coworkers\cite{Dahm2009},
however, it was strongly suggested that in underdoped
YBa$_{2}$Cu$_{3}$O$_{6.6}$ (YBCO), the kink is due to the upper branch
of the hourglass dispersion of spin fluctuations, rather than to the
resonance mode. This has opened the question of how the influence of
the resonance mode and that of the upper branch cooperate, under which
conditions the former is the dominant one, and under which the
latter.

A relevant piece of information was recently reported by
\citeauthor{Plumb2013}\cite{Plumb2013}. These authors have shown that
in nearly optimally doped Bi$_{2}$Sr$_{2}$CaCu$_{2}$O$_{8+\delta}$
(Bi2212), the energy of the kink decreases as a function of the angle
between the Fermi surface cut and the Brillouin zone diagonal, from
about \SI{65}{\milli\electronvolt} at the node (i.e., at the
diagonal), to about \SI{55}{\milli\electronvolt} roughly one-third of
the way to the antinode. In addition, when going from the node to the
antinode, the kink and also the underlying structures of the
quasiparticle self-energy sharpen dramatically. These trends of the
kink energy and sharpness have been compared with simple estimates for
several phonon modes and for the upper branch of the hourglass of spin
fluctuations, and the greatest similarity has been found for the latter.

The aims of the present study are (a) to address the angular
dependence of the kink using the fully selfconsistent version of the
Eliashberg equations employed in previous studies by some of the
authors\cite{Chaloupka2007,Sopik2015}, and the same inputs (band structure and spin
susceptibility) as in Ref.~\onlinecite{Dahm2009}, and to find out whether the model is
capable of accounting for---in addition to the nodal dispersion---the
trends reported recently by \citeauthor{Plumb2013} (b) To clarify the interplay
between the roles of the resonance mode and of the upper branch of the
hourglass in the formation of the kink.

The rest of the paper is organized as follows. In
Sec.~\ref{sec:theory} we summarize the equations employed in the
calculations, present important computational details and discuss our
choice of the values of the input parameters. Our results are
presented in Secs.~\ref{sec:results_dispersion} and
\ref{sec:off_diag}. In Subsection~\ref{sec:nodal_kink}, we address
qualitative aspects of the nodal kink, among others the role played by the
kurtosis of the resonance mode of the spin
susceptibility. In Subsection~\ref{sec:appendix}, we provide a detailed
account of the relation between the energy and the shape of the nodal
kink, and the structures of the quasiparticle self-energy. In
particular, we highlight the effect of the magnitude of the coupling
constant on the properties of the kink. In Sec.~\ref{sec:off_diag} we
address the evolution of the kink when going from the node to the
antinode. First (in Subsec.~\ref{sec:sigma_eff}), we use the effective
self-energy approach of Ref.~\onlinecite{Plumb2013} and then (in
Subsec.~\ref{sec:analysis}) our own approach based on an approximate
relation between the properties of the kink and those of the quantity
$S(\v{k},E) \equiv \Sigma_0(\v{k},E) + \phi(\v{k},E)$. Here
$\Sigma_0(\v{k},E)$ and $\phi(\v{k},E)$ are the $\tau_0$ component of
the self-energy and the anomalous self-energy, respectively. In
Sec.~\ref{sec:discussion} we compare our results with the experimental
data of Refs.~\onlinecite{Dahm2009} and \onlinecite{Plumb2013}. It is shown
that a minor modification of the input parameter values brings the
renormalized (nodal) Fermi velocity and the energy of the nodal kink
close to the experimental values for YBCO\cite{Dahm2009}. The
calculated magnitude of the slope of the angular dependence 
of the kink energy is only slightly larger than that of
Bi2212\cite{Plumb2013}. We make a prediction concerning the angular
dependence of the kink energy in underdoped YBCO and provide a
possible qualitative interpretation of the difference between the kink
in underdoped YBCO and that in Bi2212.


\section{\label{sec:theory}Spin-fermion model based calculations}

Within the spin-fermion model\cite{Moriya2000, Abanov2003,
bennemann_physics_2003, Eschrig2006, Carbotte2011, Scalapino2012}, the
self-energies $\widehat{\Sigma}_{A}(\v{k}, \text{i}E_n)$ and
$\widehat{\Sigma}_{B}(\v{k}, \text{i}E_n)$ of the antibonding and
bonding bands of a bilayer cuprate superconductor, such as Bi2212 or
YBCO, are given by\cite{Eschrig2002}:
\begin{equation}
\label{eq:selfE} \widehat{\Sigma}_{A / B} = g^2 \bigl[
\chi_{\text{SF}}^o \ast \widehat{\mathcal{G}}_{B / A} +
\chi_{\text{SF}}^e \ast \widehat{\mathcal{G}}_{A / B} \bigr].
\end{equation} Here $g$ is the coupling constant, whose dependence on
$\boldsymbol{k}$ is neglected, $\chi_{\text{SF}}^o(\boldsymbol{q},
\text{i} \omega_n)$ and $\chi_{\text{SF}}^e(\boldsymbol{q}, \text{i}
\omega_n)$ are the odd and even components of the spin susceptibility\cite{Fong2000},
respectively, and the symbol $\chi_{\text{SF}} \ast
\widehat{\mathcal{G}}$ stands for
\begin{equation}
\label{eq:convolution} \dfrac{1}{\beta N}
\sum\limits_{\boldsymbol{k'}, \text{i}E'_n}
\chi_{\text{SF}}(\boldsymbol{k} - \boldsymbol{k'}, \text{i} E_n -
\text{i} E'_n) \times \widehat{\mathcal{G}}(\boldsymbol{k'}, \text{i}
E'_n).
\end{equation} Further, $\widehat{\mathcal{G}}_{A / B}(\boldsymbol{k},
\text{i}E_n)$ are the Nambu propagators of the renormalized electronic
quasiparticles:
\begin{equation}
\label{eq:greens_function} \widehat{\mathcal{G}}_{A /
B}(\boldsymbol{k}, \text{i}E_n) = \dfrac{1}{\text{i}E_n
\widehat{\tau}_0 - (\epsilon^{A / B}_{\boldsymbol{k}} - \mu)
\widehat{\tau}_3 - \widehat{\Sigma}_{A / B} (\boldsymbol{k},
\text{i}E_n)},
\end{equation} where $\widehat{\tau}_0$ and $\widehat{\tau}_3$ are the
Pauli matrices, $\epsilon^{A}_{\boldsymbol{k}}$ and
$\epsilon^{B}_{\boldsymbol{k}}$ are the bare dispersion relations of
the two bands, and $\mu$ is the chemical potential. We have considered
only the odd channel (i.e., only the term with $\chi_{\text{SF}}^o$ in
Eq.~(\ref{eq:selfE})). This channel has been
demonstrated\cite{Eschrig2002} to be the dominant one, in particular
because $\chi_{\text{SF}}^e$ does not exhibit a pronounced resonance
mode\cite{Bourges1997a}. A broadening factor $\delta$ is used in the
analytic continuation of the propagators to the real axis ($\text{i}
E_n \rightarrow E + \text{i} \delta$), $\delta =
\text{\SI{1}{\milli\electronvolt}}$.

\begin{figure}
  \includegraphics{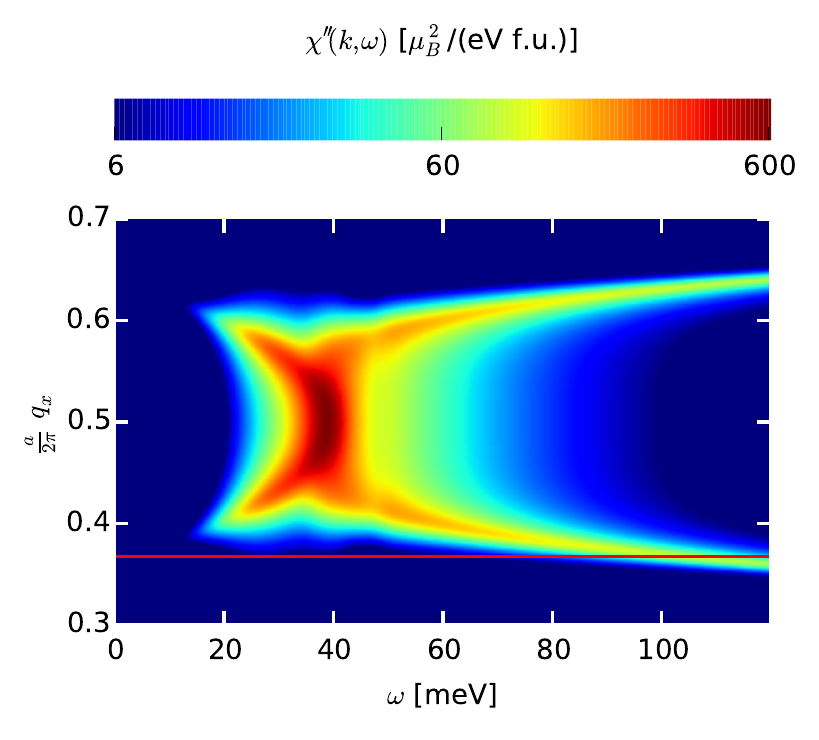}
  \caption{\label{fig:dahm_chi}
    Cut of the spin excitation spectrum $\chi''(\v{q}, \omega)$ along the nodal axis,
    calculated using the set of parameter values $S_1$. The solid red
    line corresponds to the position of the vector $\v{Q}_0$ shown in
    Fig.~\ref{fig:dahm_BZ}.}
\end{figure}

\begin{figure}
  \includegraphics{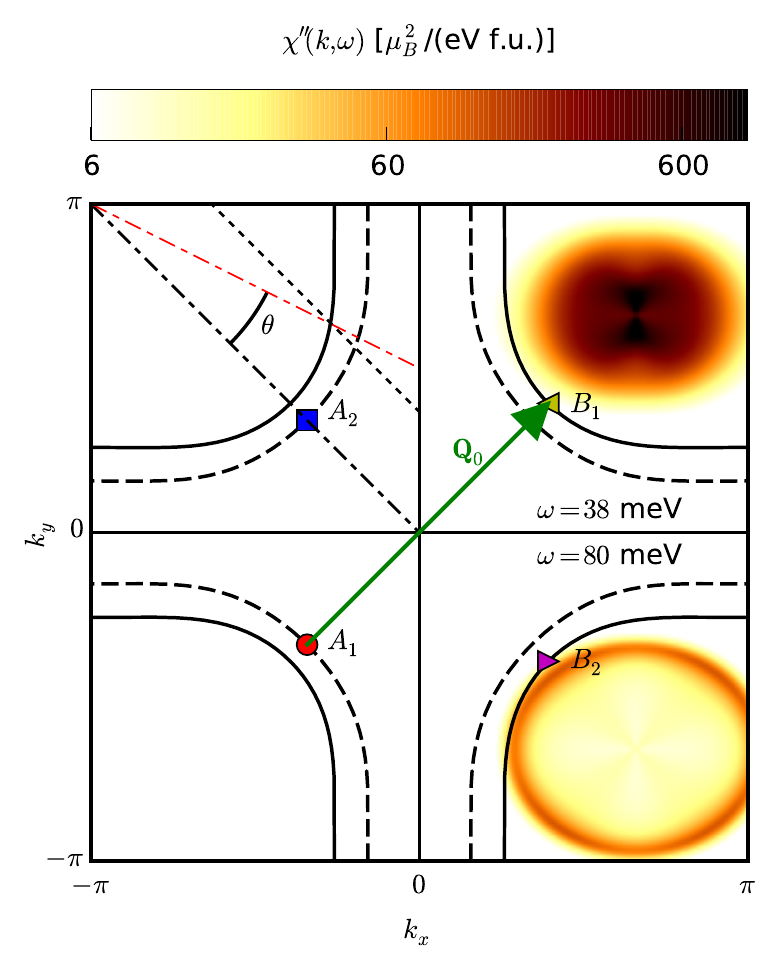}
  \caption{\label{fig:dahm_BZ} The Fermi surfaces for the antibonding
    (dashed line), and bonding (solid line) bands, obtained using the
    set of parameter values $S_1$. The solid green arrow represents
    the interband scattering vector $\v{Q_0}$. The red dashed-dotted
    line (the nearby dashed line) indicates an example of the Fermi surface cut
    used in Subsec.~\ref{sec:sigma_eff} (~\ref{sec:analysis}). Also shown are two
    (suitably shifted) constant energy cuts of the spin susceptibility. The one shown in
    the upper right quadrant corresponds to $\chi''(\v{k} -
    \v{k}_{A_{1}}, \omega=\text{\SI{38}{\milli\electronvolt}})$, the
    one shown in the bottom right quadrant to $\chi''(\v{k} -
    \v{k}_{A_{2}},
    \omega=\text{\SI{80}{\milli\electronvolt}})$.}
\end{figure}

The input parameters of the model are the imaginary component \imchi
(the indices are omitted for simplicity) of the spin susceptibility,
the dispersion relations
$\epsilon^{A/B}_{\boldsymbol{k}}$, the chemical potential $\mu$, and
the coupling constant $g$. For all of them
except for $g$, and except otherwise stated, we have used the parametrization published
in Ref.~\onlinecite{Dahm2009}, that is based on fits of the
neutron\cite{Hinkov2010} and photoemission data of underdoped
YBa$_2$Cu$_3$O$_{6.6}$. The spin susceptibility exhibits the hourglass
shape with the resonance mode at $q = (\pi / a, \pi / a)$, illustrated
in Figure~\ref{fig:dahm_chi} by a cut of the spectrum of
$\chi''(\v{q}, \omega)$ along the nodal axis. The Fermi surfaces
corresponding to the dispersion relations
$\epsilon^{A}_{\boldsymbol{k}}$ and $\epsilon^{B}_{\boldsymbol{k}}$
are shown in Fig.~\ref{fig:dahm_BZ}. The distances from the $\Gamma$
point to the Fermi surfaces, along the Brillouin zone diagonal and
expressed in units of $\dfrac{\pi}{a}\sqrt{2}$, are
$k_{F, \, N}^{A} = 0.342$, and $k_{F, \, N}^{B} = 0.393$. The
calculations are done for $T = \text{\SI{20}{\kelvin}}$.

Finally, we address the coupling constant $g$. In
Ref.~\onlinecite{Dahm2009}, the magnitude of the superconducting gap
$\Delta_{\text{SC}}$ was fixed
($\Delta_{\text{SC}}=\text{\SI{30}{\milli\electronvolt}}$), so that
the value of the coupling constant $g$ could be obtained by imposing
that the value of the calculated renormalized Fermi velocity be
consistent with the angle resolved photoemission (ARPES) data. This choice
leads to a high value of the superconducting transition temperature
$T_{\text{c}}$ of \SI{174}{\kelvin}. In the present work, the iterative solution
of Eqs.~(\ref{eq:selfE}) and~(\ref{eq:greens_function}) has
been performed in a fully self-consistent manner, along the lines of
Refs.~\onlinecite{Chaloupka2007,Sopik2015}. The renormalized
dispersions are adjusted at each iteration, following the approach developed in
Refs.~\onlinecite{Dahm2009,Dahm2013}, in such a way that the renormalized Fermi
surfaces are fixed and match the ARPES profiles used as inputs. Within this framework, $\Delta_{\text{SC}}$
is not constrained, so that its dependence on $g$ has allowed us to
fix the value of $g$ by requiring that
$\Delta_{\text{SC}}=\text{\SI{30}{\milli\electronvolt}}$. The
resulting value of $g$ of \SI{1.0}{\electronvolt} is considerably
smaller than that of Ref.~\onlinecite{Dahm2009} (the coupling constant
of the latter reference $\bar{U}$ is connected to our $g$ by $\bar{U}
= g \sqrt{{2}\over {3}}$, and the value of $\bar{U}$ used therein corresponds
to $g=\text{\SI{1.95}{\electronvolt}}$). The renormalization of the
nodal Fermi velocity is weaker and the value of $T_\text{c}$ lower
with this smaller value of $g$. The set of parameter values just
introduced is the main set used throughout the paper, and is
referred to as set $S_1$.

The calculations have been performed using the fast Fourier transform
algorithm, taking full advantage of the symmetries of the system. We
have used a grid of $256 \times 256$ points in the Brillouin zone and
a cutoff of \SI{4}{\electronvolt} to limit the number of Matsubara
frequencies. We have checked, by varying the density of the grid and
the cutoff, that these values are sufficient.

\section{\label{sec:results_dispersion}The kink in the dispersion
relation along the nodal axis}

\subsection{\label{sec:nodal_kink}Role of the upper branch of
  \imchi}

\begin{figure}
  \includegraphics{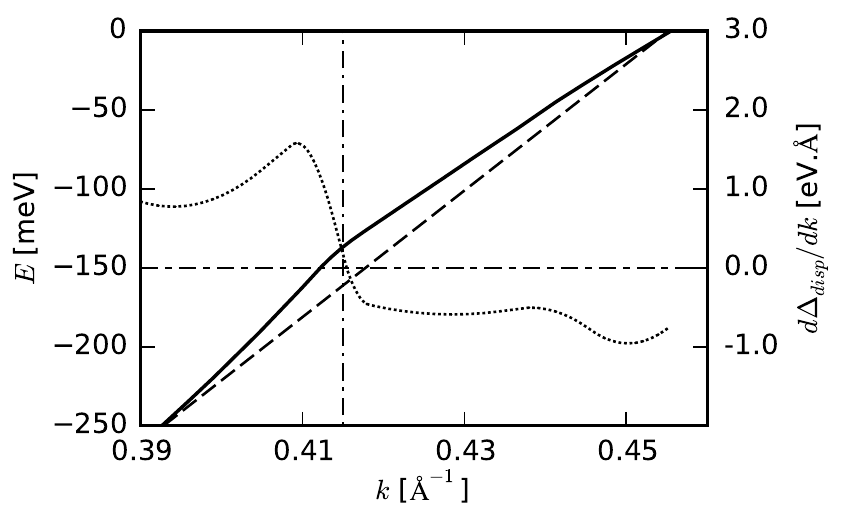}
  \caption{\label{fig:simple_nodal_kink_TD_g1} Dispersion relation
    along the Brillouin zone diagonal for the bonding band. The solid
    line represents the renormalized dispersion. The dashed line
    represents a linear approximation to the bare dispersion. The
    dotted line is the derivative of the difference
    $\Delta_\text{disp}$ between the renormalized dispersion and the
    bare dispersion. The vertical dash-dotted line is a guide to the
    eye. The calculations have been performed using the set of
    parameter values $S_1$.}
\end{figure}

The solid blue line in Fig.~\ref{fig:simple_nodal_kink_TD_g1}
represents the electronic dispersion along the nodal axis for the
bonding band. For a given energy, the associated value of $k$ is
obtained as the root of the real part of the denominator of
Eq.~(\ref{eq:greens_function}). It coincides with the value of $k$
corresponding to the maximum of the spectral function for the given
energy. The dashed line connects the quasiparticle peak at $k_F$
and the maximum of the spectral function corresponding to the high
energy cutoff of \SI{250}{\milli\electronvolt}. The kink is
smooth and broad, with a relatively small amplitude. The discrepancy
between this profile and the result of Ref.~\onlinecite{Dahm2009} is
mainly due to the lower value of $g$ used in the present study, as
discussed in detail in Subsec.~\ref{sec:appendix}.

The position and the profile of the kink can be understood in terms of
a combination of the geometrical features of the Fermi surfaces and
those of the spin susceptibility spectrum. Consider a scattering
process whereby an electron from the bonding band, of quasimomentum
$\v{k}$ and energy $E$, is scattered to the antibonding band,
quasimomentum $\v{k} - \v{q}$ and energy $E - \omega$, while a spin
excitation of quasimomentum $\v{q}$ and energy $\omega$ is emitted (an
example with $\v{k} = \v{k}_{B_{1}}$ and $\v{q} = \v{Q}_{0} \equiv
\v{k}_{B_{1}} - \v{k}_{A_{1}}$ is shown in
Fig.~\ref{fig:dahm_BZ}). The process can occur with a considerable
probability only if the momentum $\v{q}$ is such that $\chi''(\v{q},
\omega)$ is significant. Let us consider scattering processes along
the direction of the Brillouin zone diagonal, from the region around
$\v{k}_{B_{1}}$ to the region around $\v{k}_{A_{1}} = \v{k}_{B_{1}} -
\v{Q}_{0}$. Figure~\ref{fig:dahm_BZ} shows that such processes have a
negligible probability for $\omega \simeq
\text{\SI{40}{\milli\electronvolt}}$ (see the constant energy cut
shown in the upper right quadrant of Fig.~\ref{fig:dahm_BZ}). The
contribution of the resonance mode to the quasiparticle self-energy
$\widehat{\Sigma}^{B}_{|\v{k} = \v{k}_{B_1}}$ can thus be expected to
be negligible, and the nodal dispersion to be almost unaffected by the
presence of the resonance mode. For $\omega \simeq
\text{\SI{80}{\milli\electronvolt}}$ -- the energy of the crossing
point of the red line and the upper branch of the hourglass in
Fig.~\ref{fig:dahm_chi} --, however, the probability is considerable
(see the constant energy cut in the lower right quadrant of
Fig.~\ref{fig:dahm_BZ}). The nodal dispersion can thus be expected to
be strongly influenced by the coupling to spin excitations of the
upper branch. Indeed, the calculated spectrum of
$\Im{\widehat{\Sigma}^{B}_{|\v{k} = \v{k}_{B_1}}}$, shown in
Fig.~\ref{fig:selfE_base}, does not exhibit any significant feature
around \SI{40}{\milli\electronvolt} due to the resonance
mode. Instead, it displays a steep onset around
\SI{80}{\milli\electronvolt} due to the upper branch.

The kink itself (defined as the minimum of the second derivative of
the dispersion) is located at a higher energy of 
about \SI{130}{\milli\electronvolt}. The difference is due to two facts. (a)
The kink energy corresponds to the energy of the maximum of the real
part of the self-energy (connected to its imaginary part through the
Kramers-Kronig relation). This maximum is located at an energy higher
than that of the onset of the imaginary part. This issue is discussed
in detail in Subsec.~\ref{sec:appendix}. (b) The self-energy is
$k$-dependent and in the region of $k$-space around the kink (where
$|\v{k}| < k_{F, \, N}^{B}$), its imaginary part sets on at a higher
energy than for $k$ close to $k_{F, \, N}^{B}$. This can be inferred
from Figure~\ref{fig:dahm_chi}: the energy of the crossing point of
the upper branch of \imchi with a fixed $q$ horizontal line increases
when the magnitude of $q$ decreases. The impact of the $k$-dependence
of the self-energy on the energy of the kink is quantitatively
assessed in Subsec.~\ref{sec:appendix}. The validity of the simple
relation between the kink energy and the boson energy has been
examined, in a different context, by Schachinger and
Carbotte\cite{Schachinger2009}.

The above analysis confirms the conclusions of Ref.~\onlinecite{Dahm2009}
regarding the origin of the kink. However, it additionally reveals
that the presence of the upper branch {\it per se} is not a sufficient
condition for it to play the prominent role in the formation of the
nodal kink. Another necessary condition is the simultaneous occurrence
of a low kurtosis\cite{DeCarlo1997} of $\chi''(\v{q},
\omega_{\text{res}})$ (where $\omega_{\text{res}}$ is 
the frequency of the resonance mode) and of a 
relatively small value of $| \v{Q}_0 |$. Only under these
conditions is the contribution of the resonance mode negligible. A higher
kurtosis of $\chi''(\v{q}, \omega_{\text{res}})$ or a larger value of
$| \v{Q}_0 |$ would allow the contribution of the resonance to be
large enough and dominate
that of the high-energy branch. This effect was confirmed by separate
calculations of the respective contributions of the resonance mode and
of the upper branch/continuum for various shapes of the spectrum of
\imchi.

\begin{figure}
  \includegraphics{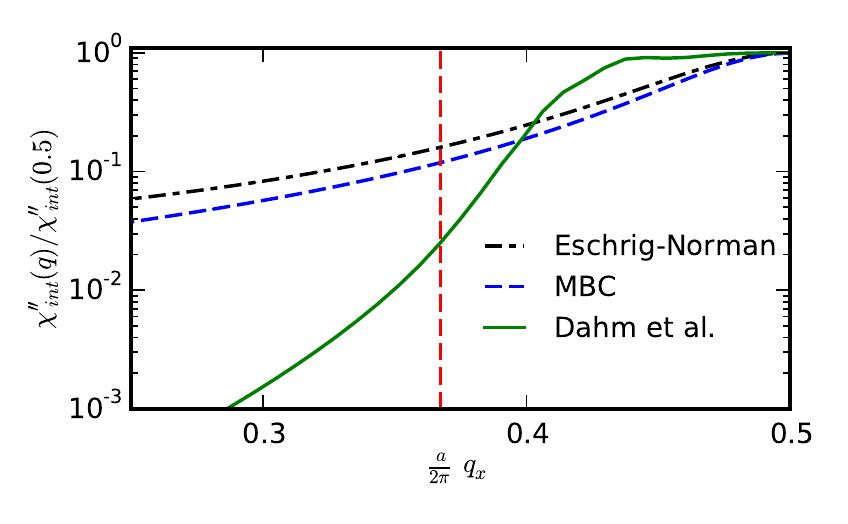}
  \caption{\label{fig:local_chi_comp}
    The quantity $\chi''_{\text{int}}$, defined in the text, as a
    function of $q_x$ along the Brillouin zone diagonal. The three
    lines correspond to the three profiles of $\chi''(\v{q}, \omega)$
    discussed in the text. The vertical red dashed line indicates the
    position of the interband vector $\v{Q_0}$.}
\end{figure}

The low kurtosis exhibited by $\chi''(\v{q}, \omega_{\text{res}})$ is
illustrated in Fig.~\ref{fig:local_chi_comp}, which displays
$\chi''_{\text{int}}(q) =
\int_0^{\text{\SI{40}{\milli\electronvolt}}}\chi''(q, \omega) d\omega$
as a function of $q$ for $\v{q}$ along the Brillouin zone
diagonal. The figure allows us to assess the $q$-space distribution of
the spectral weight of low energy spin fluctuations including the
resonance mode. The solid green line, corresponding to the spectrum of
\imchi used in the present study, exhibits a broad peak and thin
tails, both characteristic of a distribution with low kurtosis. The
dashed blue line corresponds to the form of the spin susceptibility
used by two of the present authors in previous
studies\cite{Munzar1999,Casek2005,Chaloupka2007} (the MBC form in the
following). It possesses a higher kurtosis, with both a narrower peak
and fatter tails. Finally, the black dash-dotted line represents the
susceptibility profile used by \citeauthor*{Eschrig2003} in their
thorough analysis of the dispersion anomalies within the spin-fermion
model\cite{Eschrig2003} (see also Ref.~\onlinecite{Eschrig2006}). It
also displays a relatively high kurtosis. The vertical red dashed line
sits at the position of the interband vector $\v{Q}_0$. It can be seen
that both for the MBC profile and for the Eschrig-Norman one,
$\chi''_{\text{int}}(|\v{Q}_0 |)$ is significant, approximately an
order of magnitude larger than the corresponding value for the present
spectrum of \imchi. This has a direct impact on the magnitude of the
contribution of the resonance mode to the quasiparticle
self-energy. Note, that the spectrum of \imchi used here was obtained from a fit
to experimental inelastic neutron scattering data, while the other two
spectra (MBC and Eschrig-Norman) are based on assumptions about the
$q$-dependence.
The considerations here are complementary to those of a previous work
by \citeauthor{Chubukov2004}\cite{Chubukov2004}, where the weakening
of the effect of the resonance on the near nodal dispersion has been
addressed using an analytical approach.

\subsection{\label{sec:appendix}Impact of the magnitude of the
  coupling constant}

In this subsection, we examine the link between the kink in the nodal
dispersion and the features of the fermionic self-energy. Using
Eq.~(\ref{eq:greens_function}), we find that the renormalized velocity
$v$ for a quasimomentum $k$ along the nodal axis is given
by:
\begin{equation}
\label{eq:kink_position} v(\bar{\epsilon}_k) =
\dfrac{v_0(\bar{\epsilon}_k) + \partial_{k}\Sigma'(k,
\bar{\epsilon}_k)}{1 - \partial_{E}\Sigma'(k, \bar{\epsilon}_k)},
\end{equation} where $v_0$ is the bare velocity and $\bar{\epsilon}_k$
the renormalized dispersion. The known form of the bare velocity
allows one to approximate $v_0(\bar{\epsilon}_k)$ by its value at the
Fermi surface, $v_{F_0}$.  Moreover, it is usually assumed that the
momentum dependence of the self-energy is weak\cite{Eschrig2006}, so
that the term $\partial_{k}\Sigma'(k, \bar{\epsilon}_k)$ in
Eq.~(\ref{eq:kink_position}) can be neglected, and the term
$\partial_{E}\Sigma'(k, \bar{\epsilon}_k)$ replaced with
$\partial_{E}\Sigma'(k=k_F, \bar{\epsilon}_k)$. With these
approximations, the energy dependence of $v$ is determined by the
renormalization factor $Z(\bar{\epsilon}_k) = 1
- \partial_{E}\Sigma'(k=k_F, \bar{\epsilon}_k)$, and the energy of the
kink coincides with the energy of the 
extremum of $\Sigma'(k=k_F, \bar{\epsilon}_k)$. In the following, we
quantitatively assess the impact of the momentum dependence of the
self-energy on the kink energy and shape, and identify two
qualitatively distinct regimes.

\begin{figure}
  \includegraphics{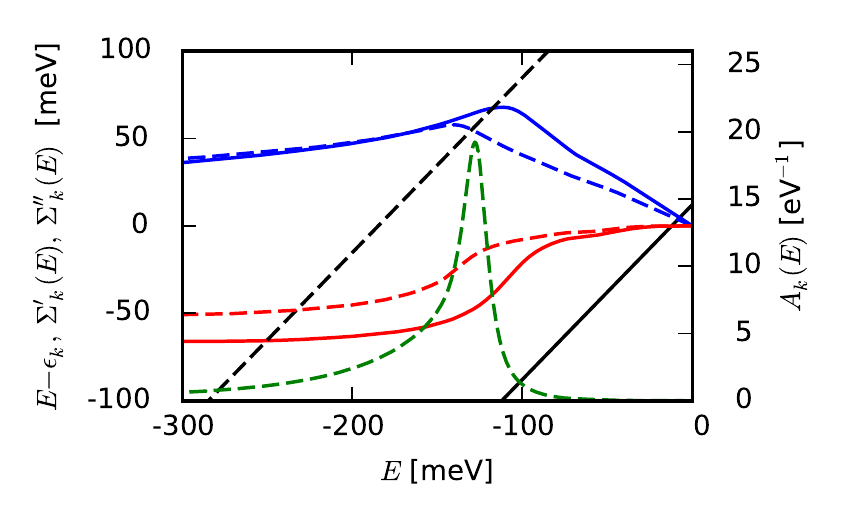}
  \caption{\label{fig:selfE_base} Graphical solution of the equation
    for the quasiparticle energy $\bar{\epsilon}_k$, for two different
    values of $k$ along the nodal axis: $k = k_{F, \, N}^{B}$ and
    $k_{\text{kink}}$ (i.e., the value of quasimomentum for which the nodal
    kink occurs), and the corresponding spectra of the real and imaginary
    parts of the self-energy, and of the spectral function
    $A_k(E)$. The calculations have been performed using the set of parameter
    values $S_1$. The solid lines
    correspond to $k = k_{F, \, N}^{B}$, the dashed lines to $k =
    k_{\text{kink}}$. The black lines represent the linear functions $E -
    \epsilon_k - \mu$, the red lines the imaginary parts of the self-energy,
    whose real parts are shown in blue. The green line represents the
    spectral function for $k_{\text{kink}}$.}
\end{figure}

Figure~\ref{fig:selfE_base} illustrates the relationship between the
energy of the kink and the energies of the features of the
self-energy, for the set of parameter values $S_1$. It shows the
graphical solution of the equation for the quasiparticle energy
$\bar{\epsilon}_k$, for two values of $k$ along the nodal axis: $k_{F,
\, N}^{B}$ and $k_{\text{kink}}$ (the value of quasimomentum at which
the kink occurs). Also shown are the corresponding spectra of the real and
imaginary components of the normal self-energy, and for
$k_{\text{kink}}$, in addition, the normal
spectral function $A_k(E)$. The spectral function for  $k_{F,
\, N}^{B}$ possesses a sharp quasiparticle peak at $E=0$. For each of the two values of $k$,
$\bar{\epsilon}_k$ is determined as the energy of the crossing between
the corresponding black line (representing $E - \epsilon_k + \mu$) and
the corresponding blue line (representing $\Re{\Sigma(k, E)}$). The
energies of the crossing points coincide with those of the
quasiparticle peaks of $A_k(E)$, as expected. It can be seen that
$\Sigma''_{k = k_{F, \, N}^{B}}$ sets on at around
\SI{80}{\milli\electronvolt} as discussed in
Sec.~\ref{sec:nodal_kink}, and that the maximum of its Kramers-Kronig
transform $\Sigma'_{k = k_{F, \, N}^{B}}$ occurs at a higher energy
(approximately \SI{110}{\milli\electronvolt}) due to the finite width
of the step in $\Sigma''_{k = k_{F, \, N}^{B}}$. Finally, the
aforementioned assumption of weak momentum dependence of the
self-energy can be seen to be valid: even though the energy of the
maximum of $\Sigma'_{k = k_\text{kink}}$ is higher than that of the
maximum of $\Sigma'_{k = k_{F, \, N}^{B}}$ by $\Delta_{\text{kink}}
\simeq \text{\SI{20}{\milli\electronvolt}}$, the shapes of the
profiles are qualitatively very similar. In particular, a sharp
maximum is present in both profiles. This explains why the energy of
the kink is only slightly (by $\Delta_{\text{kink}}$) higher than that
of the maximum of $\Sigma'_{k = k_{F, \, N}^{B}}$, and why the kink is
relatively sharp.

\begin{figure}[h]
  \includegraphics{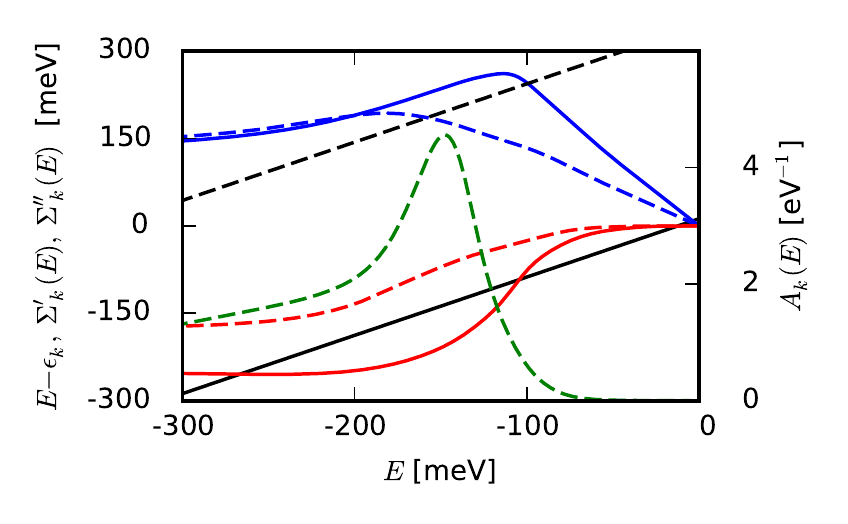}
  \caption{\label{fig:selfE_Dahm} The same quantities as in
    Fig.~\ref{fig:selfE_base}, calculated with the same input parameter
    values, except for $g=\text{\SI{1.95}{\electronvolt}}$, consistent
    with Ref.~\onlinecite{Dahm2009}. Notice the change in the scale of the
    left axis, compared with Fig.~\ref{fig:selfE_base}.}
\end{figure}

It is worth contrasting these findings with the results of the fully
self-consistent approach with the value of the
coupling constant $g$ of \SI{1.95}{\electronvolt}
(as in Ref.~\onlinecite{Dahm2009}) in place of $g =
\text{\SI{1.0}{\electronvolt}}$. Figure~\ref{fig:selfE_Dahm}
illustrates the properties of the system in this case. The large value
of the coupling constant induces much larger magnitudes of the real
and imaginary parts of the self-energy than in the former case. Thus,
the maximum value of $\Sigma'_{k=k_{F, \, N}^{B}}$ is much larger, and
the distance between $k_{F, \, N}^{B}$ and $k_\text{kink}$ as
well. Figure~\ref{fig:selfE_Dahm} shows that over such a broad
$k$-interval, the quasimomentum dependence of $\Sigma'(k, E)$ may no
longer be considered to be weak. The flattening of $\Sigma'$ as $k$
moves away from the Fermi surface (expected irrespective of the chosen
set of parameter values) is large enough for the profile to change 
qualitatively. In particular, the pronounced maximum of
$\Sigma'$ disappears before the $E - \epsilon_k + \mu$
line reaches it. Therefore, the position and the shape of this
extremum at $k_{F, \, N}^{B}$ are not the critical factors determining the energy and the
shape of the kink anymore. Instead, the dependence of the self-energy on $k$ has a
substantial impact on the profile of the kink. In terms related to
Eq.~(\ref{eq:kink_position}), this means that the weak momentum
approximation breaks down.

The interpretation of the formation of the kink therefore differs
qualitatively between the former and the latter case. In the low-$g$
regime, the energy of the kink is approximately given by the energy of
the maximum of $\Sigma'(k_{F, \, N}^{B}, \bar{\epsilon}_k)$, and the
kink is sharp. In the high-$g$ regime, the kink is made
smoother by the influence of the momentum dependence of
$\Sigma'$.

\section{\label{sec:off_diag}The kink in the dispersion relation
  away from the nodal axis}

Having analyzed the behavior of the kink in the dispersion relation
along the Brillouin zone diagonal, we now proceed to examine how the
situation evolves away from the nodal axis, as a function of the angle
$\theta$ between the direction of the Fermi surface cut and the
diagonal (for a definition of $\theta$, see
Fig.~\ref{fig:dahm_BZ}).

\subsection{\label{sec:sigma_eff}Effective self-energy approach}

\begin{figure}
  \includegraphics{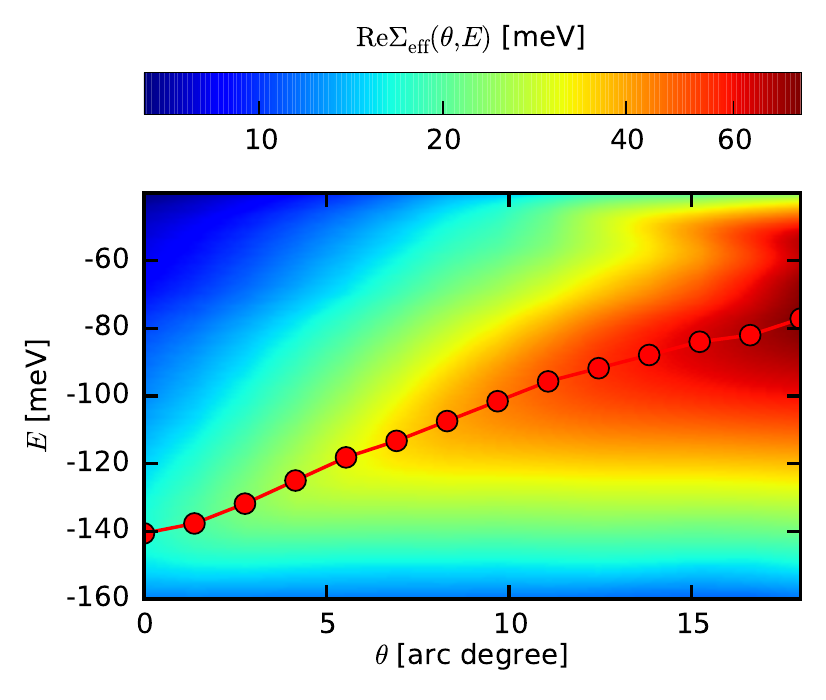}
  \caption{\label{fig:theta_kink_TD}Heat map of the real part
     of the effective self-energy $\Sigma_\text{eff}(\theta, E)$
    defined in the text, calculated using the set of parameter values
    $S_1$. For each of the selected values of $\theta$,
    the red circle represents the energy of the maximum of
    $\Re{\Sigma_{\text{eff}}(\theta, E)}$, which coincides with the energy of the
    kink.}
\end{figure}

First, we follow the approach introduced by
\citeauthor{Plumb2013}~\cite{Plumb2013}.
Figure~\ref{fig:theta_kink_TD} shows a heat map of
$\Re{\Sigma_{\text{eff}}(\theta, E)}$, the real part of the effective
self-energy defined by Eq.~(1) of Ref.~\onlinecite{Plumb2013}, and
used in order to track the angular dependence of the
kink\cite{Plumb2013}. For the convenience of the reader, the definition of
$\Sigma_{\text{eff}}(\theta, E)$ will be restated here. Denote the
inverse of the renormalized dispersion relation for a given value
of $\theta$ by $\bar{k}(\theta, E)$. Then we define
$\Re{\Sigma_{\text{eff}}(\theta, E)} \equiv \bar{\epsilon}_{k =
\bar{k}(\theta, E)} - \epsilon_{k = \bar{k}(\theta, E)}$. In
the present work, we have followed the approach of
Ref.~\onlinecite{Plumb2013}, and approximated the bare dispersion by a
straight line connecting the quasiparticle peak at $k_F$ and the 
maximum of the spectral function corresponding to the high energy cutoff of
\SI{200}{\milli\electronvolt}. The heat map has been obtained by an
interpolation of the results for a discrete set of
$\theta$-values. For each of these values, the red circle indicates
the energy of the maximum of $\Re{\Sigma_{\text{eff}}}$, coinciding with
the energy $\Omega_\text{kink}(\theta)$ of the kink in the fermionic
dispersion.

The most striking aspect of the result is the strong angular
dependence of $\Omega_\text{kink}$.  With increasing $\theta$,
$\Omega_\text{kink}$ decreases and the intensity and the sharpness of the kink
increase.  Both observations are in qualitative
agreement with the experimental findings of
Ref.~\onlinecite{Plumb2013}. These trends can be understood in terms
of the interplay between the fermionic dispersion and the bosonic
spectrum, discussed for the case of
$\theta=\text{\ang{0}}$ in Sec.~\ref{sec:nodal_kink}. As the Fermi
surface cut moves away from the nodal axis, the modulus of the
interband scattering vector along the $(\pi / a, \pi / a)$ direction
increases. As a consequence, the section of \imchi which
contributes most to the scattering, changes. As Fig.~\ref{fig:dahm_chi} shows,
the spectral weight of the constant-$q$ cut of
the upper branch of \imchi increases, and the energy of the maximum
decreases as $q$ increases towards $0.5$ from below.
The profile of the self-energy can be expected to follow the same
trend, which indeed occurs in Fig.~\ref{fig:theta_kink_TD}.

Following this analysis, we are in a position to conjecture that for
large values of $\theta$, the contribution of the resonance mode to
the scattering becomes large, and eventually dominates the
profile. This should be accompanied by a 
change of sign of the slope of $\Omega_\text{kink}(\theta)$ at a
critical angle $\theta_c$. Simple
geometrical considerations based on Fig.~\ref{fig:dahm_BZ} provide
$\theta_c \simeq \text{\ang{28}}$.
The coupling to the resonance mode has been put forward as the
source of the dispersion anomalies in earlier spin-fermion model based
studies\cite{Eschrig2006,Eschrig2003}. Within the framework of these
studies, however, the scattering mechanism does not exhibit a very
strong angular dependence, given the high kurtosis of the resonance
mode. A more precise analysis of the situation, presented in
Sec.~\ref{sec:analysis}, shows that $\theta_c$ is larger than
\ang{20}, and that for $\theta > \theta_c$, the effective self-energy
approach introduced above does not provide reliable estimates of the kink
energy.

Note finally that the scenario outlined above is -- from the
qualitative point of view -- analogous to the one proposed by
\citeauthor*{Hong2013}\cite{Hong2013}. These authors have also argued
that the observed complex structure of the quasiparticle self-energy
and its evolution when going from the nodal cut to the antinodal one
is determined by the presence of two independent contributions: that
of a resonance mode and the one of a separate branch of bosonic
excitations.

\subsection{\label{sec:analysis}Relation between the kink and the
  features of the quasiparticle self-energy}

Here we present a different approach to determine the angular
dependence of the kink energy, based on a numerical procedure for
estimating the roots of the real part of the denominator of the
Green's function~(\ref{eq:greens_function}). This method is
particularly well suited to the study of the kink for larger values of
$\theta$. For numerical reasons we use here slightly different Fermi
surface cuts than in Subsec.\ref{sec:sigma_eff}. The present ones are
parallel to the Brillouin zone diagonals. For an example of the two
types of cuts, see Fig.~\ref{fig:dahm_BZ}.

The $2 \times 2$ self-energy matrix can be expressed in
terms of the Pauli matrices:
\begin{eqnarray*} \widehat{\Sigma}(\v{k},E) \equiv \Sigma_0(\v{k},E)
  \hat{\tau_0} + \xi(\v{k},E)\hat{\tau_3} + \phi(\v{k},E)\hat{\tau_1},
\end{eqnarray*}
and the Nambu propagator as
\begin{eqnarray*} &\widehat{G}(\v{k},E) = \biggl[
\widehat{G}_0^{-1}(\v{k},E)-\widehat{\Sigma}(\v{k},E)
\biggr]^{-1} \\ =& \dfrac{\bigl[ E - \Sigma_0(\v{k},E) \bigr] \hat{\tau_0} +
\widetilde{\epsilon}(\v{k},E)\hat{\tau_3} +
\phi(\v{k},E)\hat{\tau_1}}{\bigl[ E - \Sigma_0(\v{k},E) \bigr]^2 -
\widetilde{\epsilon}(\v{k},E)^2 - \phi(\v{k},E)^{2}}.
\label{Eq:poles}
\end{eqnarray*}
We have dropped the band index for simplicity, and
$\widetilde{\epsilon}(\v{k},E)$ stands for $\epsilon(\v{k},E) - \mu +
\xi(\v{k},E)$. The normal component of the propagator is given by
\begin{equation}
  \label{eq:master2} G(\v{k},E) = \dfrac{E-\Sigma_0(\v{k},E) +
\widetilde{\epsilon}(\v{k},E)}{\bigl[ E-\Sigma_0(\v{k},E) \bigr]^2 -
\widetilde{\epsilon}(\v{k},E)^2 - \phi(\v{k},E)^{2}}.
\end{equation}

\begin{figure}
  \includegraphics{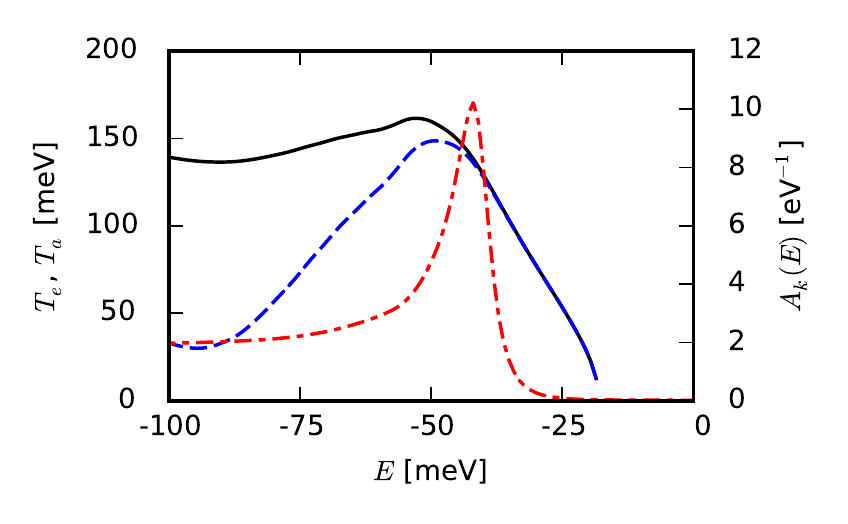}
  \caption{\label{fig:square_approx}Comparison of the expression  from
    Eq.~(\ref{eq:exact_root}), 
    $T_{\text{e}} \equiv \bigl( \Re{\left[
        \left( E - S(\v{k},E) \right) \left( E -
          D(\v{k},E) \right) \right]} \bigr) ^{1/2}$  (black solid line)  with its
    approximation $T_{\text{a}} \equiv  \bigl( \Re{\left[ E -
        S(\v{k},E) \right]} \Re{\left[E - D(\v{k},E) \right]} \bigr)
    ^{1/2}$ used in Eq.~(\ref{eq:root}) (dashed blue line), for $k = k_{\text{kink}}$
    corresponding to the cut defined by $\theta =\text{\ang{26.9}}$. The
    dashed-dotted line represents the spectral function $A_k(E)$. The
    calculations have been performed using the set of parameter values
    $S_1$.}
\end{figure}

The approach we introduce here is most easily pictured as an extension
of Sec.~\ref{sec:appendix} and Fig.~\ref{fig:selfE_base} to the case
where $\phi(\v{k},E)$ is finite. Provided the quasiparticle is well
defined, its energy $E$ is equal to the root of the real part of the
denominator, i.e., to the solution of the following equation in $E$,
parametrized by $\v{k}$:
\begin{equation}
\label{eq:exact_root} \Re{\left[ \left( E-S(\v{k},E)\right)\left(
E-D(\v{k},E)\right) - \widetilde{\epsilon}(\v{k},E)^2 \right]} = 0,
\end{equation} where $S(\v{k},E) \equiv \Sigma_0(\v{k},E) \pm
\phi(\v{k},E) $ and $D(\v{k},E) \equiv \Sigma_0(\v{k},E) \mp
\phi(\v{k},E) $. The upper (lower) sign is used if
$\Re{\Sigma_0(\v{k},E)}$ and $\Re{\phi(\v{k},E)}$ have the same
(opposite) signs (recall that $\Re{\phi(\v{k},E)}$ possesses d-wave symmetry,
while $\Re{\Sigma_0(\v{k},E)}$ is positive in the momentum-energy
section we are considering). Assuming that the imaginary parts of $E -
S(\v{k},E)$ and $E - D(\v{k},E)$ are small compared to their real
parts, we may approximate Eq.(\ref{eq:exact_root}) by:
\begin{equation}
\label{eq:root} \Re{\left[ \left( E - S(\v{k},E) \right) \right] }
\simeq \dfrac{ \Re{\left[ \widetilde{\epsilon}(\v{k},E)^2 \right]} } {
\Re{\left[ \left(E - D(\v{k},E) \right) \right] } }.
\end{equation} The validity of this assumption is related to that of
the quasiparticle picture, for an illustration, see
Fig.~\ref{fig:square_approx}.

For $\theta = \text{\ang{0}}$, $S(\v{k},E) = D(\v{k},E) =
\Sigma_0(\v{k},E)$ and Eq.~(\ref{eq:root}) reduces to the simple
equation determining the quasiparticle energy employed in
Sec.~\ref{sec:results_dispersion}, $\Re{\left[ E - \Sigma_0(\v{k},E) -
\widetilde{\epsilon}(\v{k},E) \right]} = 0$.

\begin{figure}
  \includegraphics{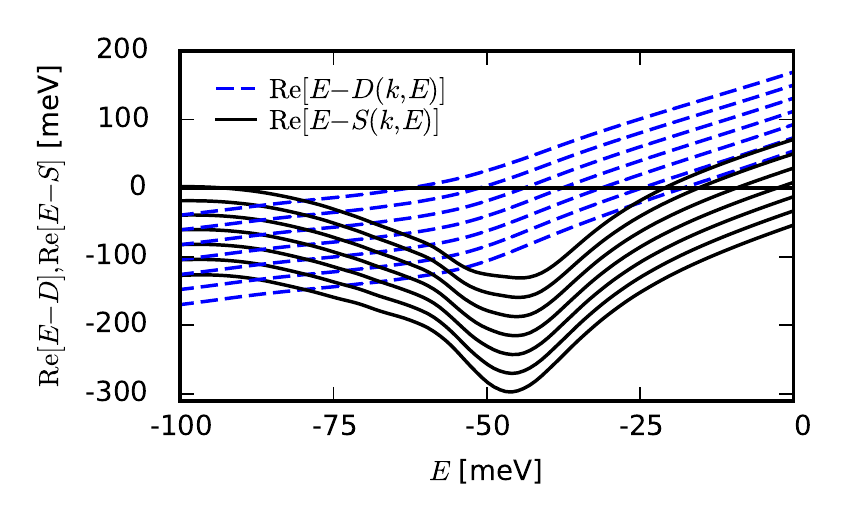}
  \caption{\label{fig:k_dependence}Profiles of the terms $\Re{\left[ E
- D(k, E) \right] }$ and $\Re{\left[ E - S(k, E) \right] }$ entering
Eq.~(\ref{eq:root}), for $\theta = \text{\ang{26.9}}$ and for a set of
values of the quasimomentum $k$, calculated using the set of parameter
values $S_1$. The lowest curves correspond to the Fermi surface. The
quasimomentum $k$ differs by $\Delta k = \pi / 128$ 
from one curve to the next. For readability, each curve is shifted by
\SI{20}{\milli\electronvolt} with respect to the previous one as $k$
moves away from the Fermi surface.}
\end{figure}

For large values of $\theta$, where the gap is fully
developed, $\Sigma_0(\v{k},E)$ and $\phi(\v{k},E)$ have comparable
magnitudes. As a consequence, $\Re{\left[ E - S(\v{k},E) \right]}$ and
$\Re{\left[ E - D(\v{k},E) \right] }$ exhibit very different
profiles, while both remain weakly $k$-dependent along a fixed
cut. This is illustrated by Fig.~\ref{fig:k_dependence}, which shows
the approximately linear profile of $\Re{\left[ E - D(k, E)
  \right]}_{|\theta =\text{\ang{26.9}}}$, contrasting with the peaked shape
of $\Re{\left[ E - S(k, E) \right] }_{|\theta=\text{\ang{26.9}}}$. The former
profile, close to linear, emerges as the difference between two
similarly peaked functions $\Sigma_0(\v{k},E)$ and $\phi(\v{k},E)$
(plus the linear function $E$). The similarity is due to the fact that
both functions result from the convolution in
Eq.~(\ref{eq:convolution}). The latter profile represents the sum of
the two functions (plus the linear function $E$), and therefore
exhibits a peaked shape reminiscent of the similar shape of both
functions.

\begin{figure}
  \includegraphics{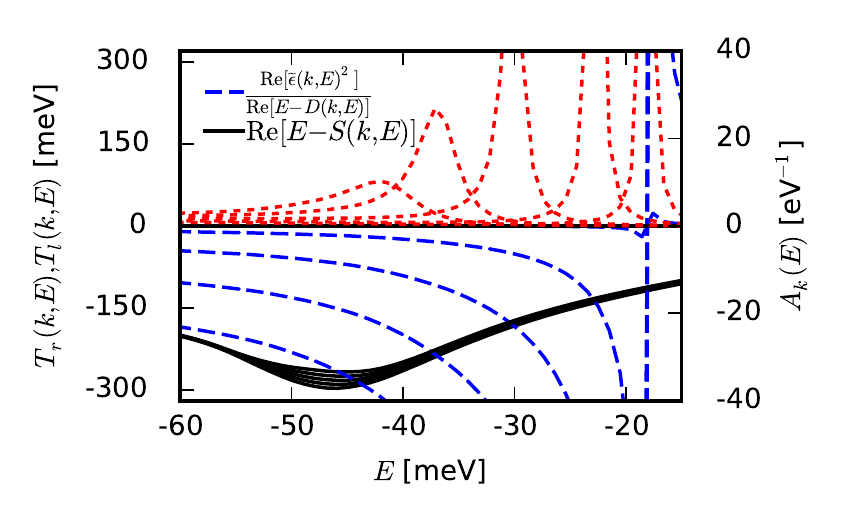}
  \caption{\label{fig:no_kink}Profiles of both sides of
    Eq.~(\ref{eq:root}) and of the quasiparticle spectral function
    $A_k(E)$ for $\theta = \text{\ang{26.9}}$ and for a set of
    values of the quasimomentum $k$, calculated using the set of parameter
    values $S_1$. As in Fig.~\ref{fig:k_dependence}, the quasimomentum $k$
    differs by $\Delta k = \pi / 128$ from one curve to the next. The set
    of dashed blue (solid black) lines represents the term $ T_r(\v{k},E) 
    \equiv \Re{\left[ \widetilde{\epsilon}(\v{k},E)^2 \right]} / \Re{\left[ \left(E -
          D(\v{k},E) \right) \right] }$ (the term $T_l(\v{k},E)  \equiv \Re{\left[ \left(E -
          S(\v{k},E) \right) \right] }$). Note that the energies of the peaks of
    the spectral function (dotted red line) coincide with those of the
    crossing points of  the corresponding blue and black lines.}
\end{figure}

The expressions entering Eq.~(\ref{eq:root}) can be interpreted in
simple terms. The one on the left hand side displays a peak whose
magnitude increases with increasing $\theta$ as a consequence of the
lengthening of the interband scattering vector, and of the 
corresponding increase of the spectral weight of the section of \imchi
which contributes to the scattering processes. The term on the
right-hand side of Eq.~(\ref{eq:root}) involves the
inverse of an approximately linear expression. For fixed values of 
$\theta$ and $\v{k}$, the value of this expression at the origin
equals $| \Re{\phi(\v{k},E=0)} |$. These observations
allow us to interpret the profile of the right-hand side of
Eq.~(\ref{eq:root}) as that of a hyperbola-like function, with the
origin of the $E$-axis displaced by $-| \Re{\phi(\v{k},E=0)} | \simeq
- | \Re{\phi(\v{k} = \v{k}_F(\theta), E=0) |} =
\Delta_{\text{SC}}(\theta)$, as illustrated in
Fig.~\ref{fig:no_kink}. As $\v{k}$ moves away from the Fermi surface, a 
family of hyperbola-like functions (``hyperbolas'' in the following)
is generated, with a multiplicative factor 
$\Re{\left[ \widetilde{\epsilon}(\v{k},E)^2 \right]}$ applied to the
$y$-axis. The right-hand side of Eq.~(\ref{eq:root}) thus evolves from
a very sharp hyperbola, for $k \to k_F(\theta)$ , to a smooth
hyperbola, for large values of $| k - k_F(\theta) |$.

This analysis shows that the left-hand (right-hand) side term of
Eq.~(\ref{eq:root}), indexed by $(k, \theta)$, is strongly (weakly)
dependent on $\theta$, but weakly (strongly) dependent on $k$. In
other words, Eq.~(\ref{eq:root}) allows us to disentangle the
sensitivities of the quantities of interest with respect to $k$ and
$\theta$. At this point, noticing that neither $\Re{
\widetilde{\epsilon}(\v{k},E)}$ nor $D(\v{k},E)$ exhibit a pronounced
kink, we are in a position to conclude that the origin of the kink in
the fermionic dispersion lies in the kink exhibited by the left hand
side of Eq.~\ref{eq:root}, $\Re{\left[ \left( E - S(\v{k},E) \right)
\right]}$. The position of the kink can now be reliably evaluated by
exploring the smooth quantity $\Re{\left[ \left( E - S(\v{k},E) \right)
\right]}$ defined on the fine energy mesh.

\begin{figure}
  \includegraphics{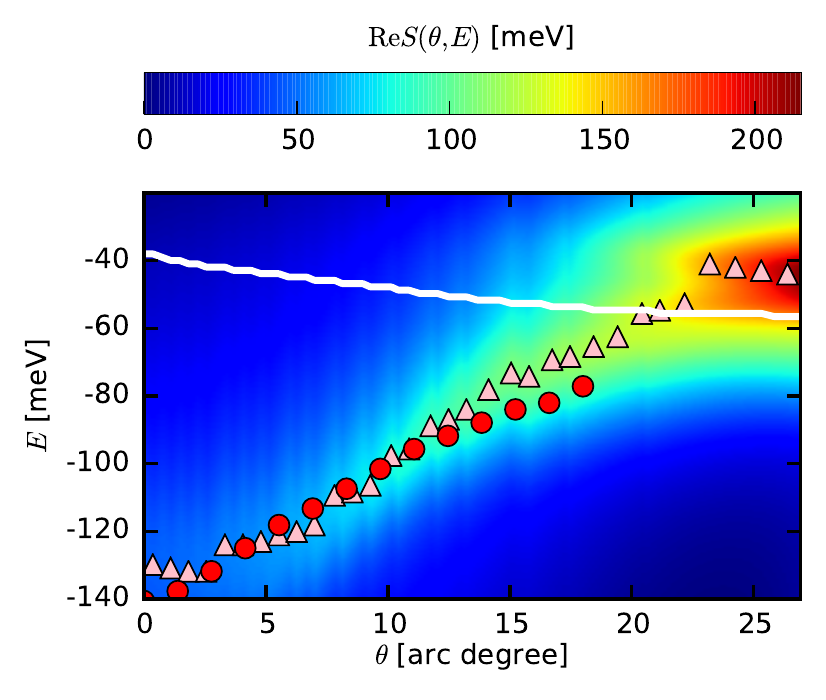}
  \caption{\label{fig:heatmap_kink}Heat map of the real part of
    the quantity $S(\v{k}, E)$ defined in the text, calculated using
    the set of parameter values $S_1$. For each of the selected values 
    of $\theta$, the pink triangle represents the energy of the extremum of
    $\Re{S(\v{k}, E)}$, which coincides with
    $\Omega_{\text{kink}}$, as discussed in the text. The solid white
    line represents the expression $\omega_{\text{res}} +
    \Delta_\text{SC}(\theta)$. The solid red circles, displayed for
    comparison, are taken from Fig.~\ref{fig:theta_kink_TD}.}
\end{figure}

The approach detailed below has been used to obtain the profile of
$\Omega_{\text{kink}}(\theta)$ displayed in
Fig.~\ref{fig:heatmap_kink}: For each selected value of $\theta$, the
momentum dependence of the self-energy is examined. We then define
$k_0 (\theta)$ as the value of $k$ on the computational $k$-mesh, along the
considered $\theta$-cut (recall that the $k$-space cuts we use in this
subsection have the advantage of matching the geometry of the
computational $k$-mesh), which is closest to $k_{\text{kink}}
(\theta)$. This process is illustrated in
Fig.~\ref{fig:no_kink}. Given a value of $\theta$, $k_0 (\theta)$ is the
value of $k$, such that the dashed line representing $ \Re{\left[
    \widetilde{\epsilon}(\v{k},E)^2 \right]} / \Re{\left[ \left(E -
      D(\v{k},E) \right) \right] }$ crosses the solid line
representing $\Re{\left[ \left( E - S(\v{k},E) \right)
\right]}$ close to its extremum. Once $k_0$ is fixed, we obtain the
energy of the kink as that of the extremum of $\Re{S(\v{k}_0,E)}$ (we have
checked that in the present context the two energies coincide). As discussed above, in the
$\theta \to 0$ limit, this method for estimating the energy of the
kink is equivalent to the one used in Sec.~\ref{sec:appendix} , but there is
one caveat: for small values of $\theta$, the gap is small, so that the
kink in $E - S(\v{k},E)$ is weak and may not always dominate the very
weak kink in $E - D(\v{k},E)$. As a consequence, for small values of
$\theta$, the former method may be more accurate in estimating the
energy of the kink.

It can be seen in Fig.~\ref{fig:heatmap_kink} that the present
$\Omega_{\text{kink}}(\theta)$ is close to the result shown in
Sec.~\ref{sec:sigma_eff}. The main discrepancies appear in the $\theta
\to 0$ region (discussed above), and for large values of $\theta$. The
latter arise because the kink becomes so intense, and sharp in
momentum space, that the former method, based on interpolations of the
renormalized dispersion in $k$-space, does not provide a precise
estimate of the kink energy.

The increased extent of the accessible $\theta$-domain allows for a
confirmation of the conjecture exposed in Sec.~\ref{sec:sigma_eff},
related to the role of the resonance
mode. Figure~\ref{fig:heatmap_kink} clearly shows that the slope of
$\Omega_{\text{kink}}(\theta)$ changes sign at $\theta_c \simeq
\text{\ang{23}}$. We argued in Sec.~\ref{sec:sigma_eff} that if the
kink is due to the upper branch of \imchi, then the slope of
$\Omega_{\text{kink}}(\theta)$ must be negative. This is the trend
observed for $\theta < \theta_c$. Conversely, if the resonance mode is
the dominant source of scattering, the $\theta$-dependence of
$\Omega_{\text{kink}}(\theta)$ is determined mainly by that of
$\Delta_{\text{SC}}(\theta)$ and $\Omega_{\text{kink}}(\theta)$ must
therefore display a positive slope close to that of
$\Delta_{\text{SC}}(\theta)$. This is what we observe in the $\theta >
\theta_c$ region of Fig.~\ref{fig:heatmap_kink}, where the profile of
$\Omega_{\text{kink}}(\theta)$ follows that of $\omega_{\text{res}} +
\Delta_{\text{SC}}(\theta)$, represented by the solid white line. The
fact that the $\Omega_{\text{kink}}(\theta)$ line is located somewhat
above the $\omega_{\text{res}} + \Delta_{\text{SC}}(\theta)$ line is
likely due to the influence of the lower branch of \imchi. The
discontinuity of $\Omega_{\text{kink}}(\theta)$ at $\theta = \theta_c$
is an artifact related to the method for the numerical determination
of $\Omega_{\text{kink}}(\theta)$.

Finally, we note the remarkable similarity between the background of
the heat map shown in Fig.~\ref{fig:heatmap_kink} and the profile of
the upper branch of \imchi displayed in Fig.~\ref{fig:dahm_chi},
arising from the selfenergy-\imchi relation~(\ref{eq:selfE}). It
illustrates the major role played by the upper branch of \imchi in the
formation of the angular dependence of $\Omega_{\text{kink}}(\theta)$
in the near nodal region.

\section{\label{sec:discussion}Comparison with experimental data}

The main trend of Subsection~\ref{sec:sigma_eff}, i.e.\ the decrease
of $| \omegakink |$ when going from the nodal cut to the antinodal
one, is consistent with the experimental findings of
Ref.~\onlinecite{Plumb2013}. Our results provide support for the
conjecture that the decrease is associated with the dispersion of the
upper branch of the hourglass. The calculated value of the energy of
the nodal kink ($\simeq$ \SI{130}{\milli\electronvolt}), however, is
much higher than that of underdoped YBCO reported in
Ref.~\onlinecite{Dahm2009} (\SI{80}{\milli\electronvolt}). In
addition, the calculated magnitude of the slope of
$\Omega_\text{kink}(\theta)$
(\SI{3.5}{\milli\electronvolt} per arc degree) is much higher than the
experimental value of Bi2212 reported in Ref.~\onlinecite{Plumb2013}
(\SI{0.8}{\milli\electronvolt} per arc degree). Finally,
the renormalized Fermi velocity of \SI{2.8}{\electronvolt\AA} on the
nodal axis (see Fig.~\ref{fig:simple_nodal_kink_TD_g1}), is much
larger than the experimental value of underdoped YBCO of
\SI{1.8}{\electronvolt\AA}. This discrepancy is connected with the
fact that the value of $g$ used in the set $S_1$ is much smaller than
that of Ref.~\onlinecite{Dahm2009}.

Based on our interpretation of the origin of the kink, it is possible
to understand the influence of the model parameters on the profile of
$\Omega_\text{kink}(\theta)$. We are also well equipped to find out
which adjustments are necessary in order to reconcile the results of
the calculations with the experimental data. It can be expected that
$\Omega_\text{kink}(\theta = 0)$ decreases with increasing interband
distance $| \v{Q}_0 |$ (see Fig.~\ref{fig:dahm_BZ} for a definition),
but that it is not very sensitive to the doping level or the
bonding-antibonding splitting (provided that $|\v{Q}_0|$ and the Fermi
velocity are kept fixed). Our analysis also indicates that a widening
of the upper branch of the hourglass should lead to a shift of
$\Omega_\text{kink}(\theta=0)$ towards lower energies and to a
reduction of the slope of $\Omega_\text{kink}(\theta)$. Finally,
reducing the bandwidth of the bare dispersion should induce a lowering
of the renormalized Fermi velocity. We have checked these trends by
performing calculations of the same type as described in sections
\ref{sec:results_dispersion} and \ref{sec:off_diag} for many different
sets of values of the input parameters.

As an example, and an illustration of the sensitivity of the results
of the calculations to the input parameter values, we present below
results of our calculations obtained using a set of parameter values
($S_2$ in the following), where some of the values have been modified
along the lines of the previous paragraph. The values of $k_{F, \,
N}^{A}$ and $k_{F, \, N}^{B}$ are increased to $36.0 \%$ and $40.7 \%$
of $\pi \sqrt{2} / a$, respectively\cite{Empty2}. This shift applied
to the band structure leaves the system well within the limits given
by published experimental values: the values of $k_{F, \, N}^{A}$ and
$k_{F, \, N}^{B}$ remain smaller than 41\%, the common value of the
two parameters reported in
Ref.~\onlinecite{Fournier2010}. Furthermore, the corresponding
increase in the magnitude of $| \v{Q}_0 |$ is small, so that the
resonance mode does not participate in the scattering along the nodal
cut, and the qualitative features of Fig.~\ref{fig:dahm_BZ} are
conserved. The bandwidth of the bare dispersion is reduced by 40\%, so
that the value of the renormalized Fermi velocity is close to the
experimental one, and we set $g=\text{\SI{0.8}{\electronvolt}}$, so
that the maximum value of the gap remains unchanged at
\SI{30}{\milli\electronvolt}. Finally, the upper branch of \imchi is
made wider, so as to further reduce the value of
$\Omega_\text{kink}(\theta = 0)$ and the slope of the profile of
$\Omega_\text{kink}$\cite{Empty1}.

\begin{figure}
  \includegraphics{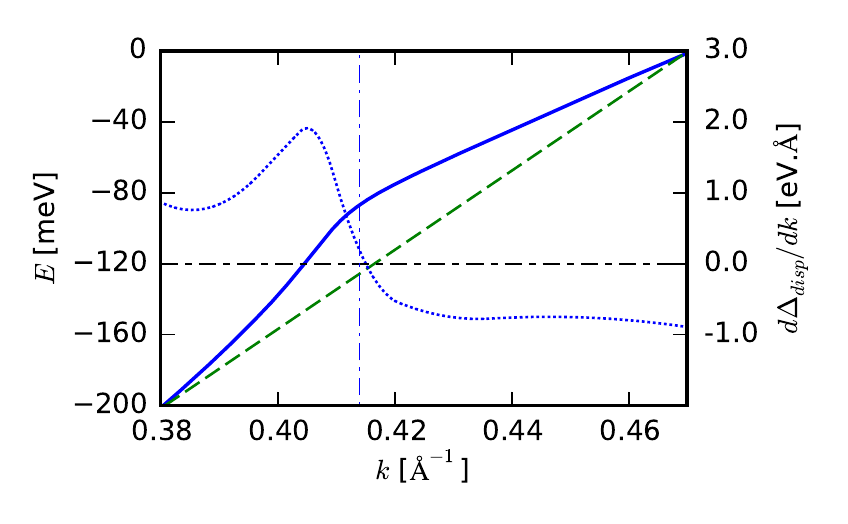}
  \caption{\label{fig:nodal_kink_TD_enhanced} The same quantities as in
    Fig.~\ref{fig:simple_nodal_kink_TD_g1}. The calculations have
    been performed using the set of parameter values $S_2$.}
\end{figure}

Figure~\ref{fig:nodal_kink_TD_enhanced} displays the renormalized
dispersion calculated using the set of parameter values $S_2$. It can
be seen that the kink is much more pronounced. As expected, the energy
of the kink (ca \SI{90}{\milli\electronvolt}) and the renormalized
Fermi velocity (ca \SI{1.5}{\electronvolt\AA}) are considerably lower
than in Fig.~\ref{fig:simple_nodal_kink_TD_g1}, and close to the
experimental values of Ref.~\onlinecite{Dahm2009}.

\begin{figure}
  \includegraphics{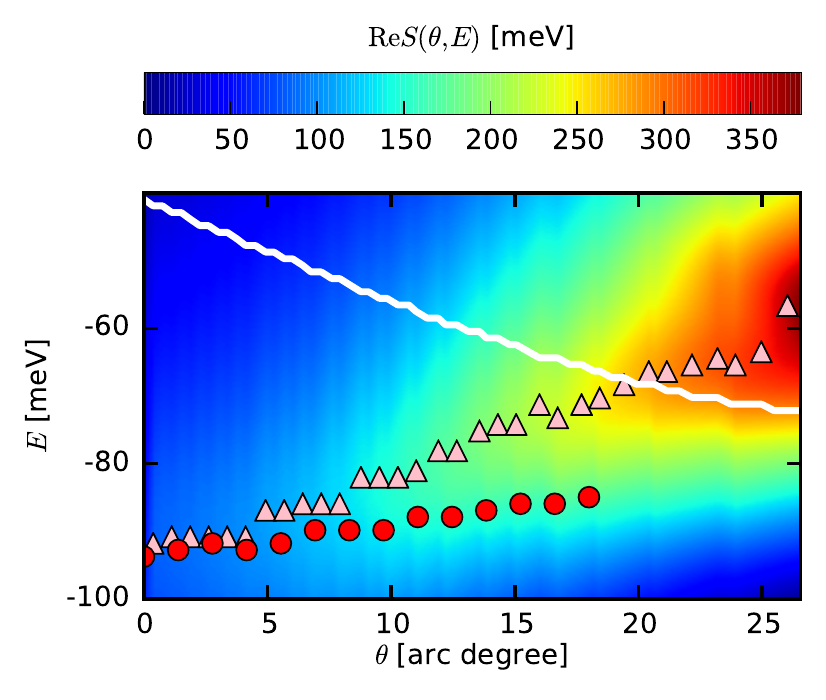}
  \caption{\label{fig:theta_kink_TD_enhanced} The same quantities as in
    Fig.~\ref{fig:heatmap_kink}, calculated using the set of input
    parameter values $S_2$. The apparent steps in the pink triangle
    profile are due to the reduced energy range of the $E$-axis, and
    the discretization of the energy mesh.}
\end{figure}

The corresponding angular dependence of \omegakink is shown in
Fig.~\ref{fig:theta_kink_TD_enhanced}. It can be seen that the
magnitude of the slope of \omegakink is reduced to only
\SI{1.1}{\milli\electronvolt} per arc degree, reasonably close to the
experimental value for Bi2212\cite{Plumb2013}. The value of $\theta_c$
of Fig.~\ref{fig:theta_kink_TD_enhanced} (ca \ang{26}) is higher than
that of Fig.~\ref{fig:heatmap_kink}. The difference is mainly due to
that between the bare dispersion relations of $S_1$ and those of
$S_2$. The interpretation exposed at the end of
Sec.~\ref{sec:off_diag} still applies. Based on this interpretation
and the above discussion we can make a prediction concerning the
angular dependence of $\Omega_{\text{kink}}$ in underdoped YBCO. We
predict that there exists a critical value $\theta_c$, such that for
$\theta < \theta_c$ ($\theta > \theta_c$),
$\Omega_{\text{kink}}(\theta)$ is a decreasing (weakly increasing)
function. The minimum $\Omega_{\text{kink}}(\theta_c)$ of
$\Omega_{\text{kink}}$ is determined by $\Delta_{\text{SC}}(\theta_c)$
and by the lower branch of \imchi. A value in the range from
\SI{40}{\milli\electronvolt} to \SI{60}{\milli\electronvolt} can be
expected. This prediction could be tested in ARPES experiments.

Finally we address, in light of our findings, the
$\Omega_{\text{kink}}(\theta)$ line for nearly optimally doped Bi2212
reported in Ref.~\onlinecite{Plumb2013}, which was one of our starting
points. The energy of the nodal kink in Bi2212 of ca
\SI{65}{\milli\electronvolt} is roughly \SI{15}{\milli\electronvolt}
lower than that of underdoped YBCO and \SI{25}{\milli\electronvolt}
lower than our result shown in
Fig.~\ref{fig:theta_kink_TD_enhanced}. The magnitude of the slope of
$\Omega_{\text{kink}}$ in Bi2212 is only slightly smaller than that of
our calculations. The difference may be caused by a difference in the
Fermi surfaces and/or by a difference in \imchi. Since the magnitude
of the internodal distance, $| \v{Q}_0 |$, of optimally doped Bi2212
is almost the same as that of underdoped YBCO, it appears that some
difference in \imchi plays the crucial role. Note that the neutron
scattering data of optimally doped Bi2212\cite{Xu2009} reveal a fairly
high kurtosis of $\chi''(q, E)_{|E =
\text{\SI{42}{\milli\electronvolt}}}$ [see Fig. 2 (c) of
Ref.~\onlinecite{Xu2009}], and that the higher energy cuts of
$\chi''(q, E)$ shown in Figs.~2(a) and~2(b) of
Ref.~\onlinecite{Xu2009} are considerably wider than those of
underdoped YBCO. In particular, the values of \imchi for $q = 0.19$
r.l.u.  (corresponding to $| \v{Q}_0 |$ of Fig.~\ref{fig:dahm_chi})
and $\omega = \text{\SI{42}{\milli\electronvolt}}$,
\SI{54}{\milli\electronvolt} and \SI{66}{\milli\electronvolt} in
Figs.~2 (c), (b) and (a) of Ref.~\onlinecite{Xu2009}, are all
significant, and of a comparable magnitude. Motivated by this
observation and by the large width of the nodal kink in Bi2212 (see
Fig. 1 (d) of Ref.~\onlinecite{Plumb2013}), we propose the following
qualitative interpretation of the angular dependence of
$\Omega_{\text{kink}}$ in Bi2212: we suggest that the nodal kink is
not determined by a single narrow cut through the upper branch of the
hourglass, as in the case of underdoped Y-123 (see Fig. 1), but rather
by a broad band of \imchi ranging from ca \SI{40}{\milli\electronvolt}
to ca \SI{100}{\milli\electronvolt}. Even the
\SI{42}{\milli\electronvolt} cut contributes because of the high
kurtosis. With increasing $\theta$, lower energy segments of \imchi
become more influential, for the same reasons as discussed in
Sec.~\ref{sec:sigma_eff}, and as a consequence, the energy of the kink
slighly decreases.

\section{\label{sec:conclusion}Summary and conclusions}

We have investigated the effect of the upper branch of the hour-glass
magnetic spectrum on the electronic dispersion of high-T$_\text{c}$
cuprate superconductors using the fully self-consistent version of the
phenomenological model, where charged planar quasiparticles are
coupled to spin fluctuations. The same input band structure and the
same input spin susceptibility as in the previous study by T. Dahm and
coworkers have been used.

First, we have confirmed the finding by \citeauthor{Dahm2009}, that
the nodal kink is determined, for the present values of the input
parameters, by the upper branch of \imchi. We have further
demonstrated that the position and the shape of the kink depend
strongly on the strength of the charge-spin coupling. For low (but
still realistic) values of the coupling constant, the position of the
kink can be estimated using the common approximation, where the
quasimomentum dependence of the self-energy along the Fermi surface
cut is neglected. The kink is weak but sharp. For high values of the
coupling constant, however, the dependence of the self-energy on the
quasimomentum plays an important role. The kink is less sharp, but has
a larger amplitude.

Second, we have shown that the kurtosis of the resonance mode of the
spin susceptibility in the quasimomentum space has a major influence
on the mechanism of the fermionic scattering. If the kurtosis is low
(high), as in the present study (as in several previous
studies\cite{Munzar1999,Eschrig2003, Casek2005,Chaloupka2007}), the
effect of the resonance mode in the near-nodal region of the Brillouin
zone is weak (large), and the upper branch of the hour-glass (the
resonance mode) plays the major role in the formation of the nodal
kink.

Third, the calculated energy of the kink decreases as a function of
the angle $\theta$ between the Fermi surface cut and the nodal
direction. This result is in qualitative agreement with recent
experimental results\cite{Plumb2013, He2013}. Based on our
interpretation of the formation of the kink, we have been able to
modify the values of the input parameters in such a way that both the
renormalized (nodal) Fermi velocity and the energy of the nodal kink
are close to the experimental values for underdoped YBCO reported by
Dahm and coworkers. The calculated magnitude of the slope of the
angular dependence of the kink energy is close to that of Bi2212
reported by Plumb and coworkers. We predict that there exists a
critical value $\theta_c$ such that the energy of the kink is a
decreasing (weakly increasing) function of $\theta$ for $\theta <
\theta_c$ ($\theta > \theta_c$) and provide a possible qualitative
interpretation of the difference between the kink in underdoped YBCO
and that in optimally doped Bi2212.

\section*{Acknowledgements}

 A part of the work at Masaryk University was carried out under the
project CEITEC 2020 (LQ1601) with financial support from the Ministry
of Education, Youth and Sports of the Czech Republic under the
National Sustainability Programme II. D.~G.~and D.~M.~were supported
by the projects MUNI/A/1496/2014 and MUNI/A/1388/2015. J.~Ch.~was
supported by the AvH Foundation and by the EC 7$^\text{th}$ Framework
Programme (286154/SYLICA). 

\recordnotes

\end{document}